\documentclass[pre,twocolumn, superscriptaddress, showpacs,amsmath, nofootinbib, floatfix] {revtex4}
\usepackage{graphicx} 
\usepackage{graphics} 
\usepackage{psfrag} 
\usepackage{epsf} 
\newcommand{\td}{\mathrm{d}}
\newcommand{\nd}{\mbox{\boldmath $n$}}
\newcommand{\nsd}{\mbox{\scriptsize\boldmath ${n}$}}
\newcommand{\ud}{\mbox{\boldmath $u$}}
\newcommand{\pr}{\left.{\hspace{-17pt}\phantom{\sum}}\right.^\prime}

\begin{document}
 
\title{Statistics of extremal intensities for Gaussian interfaces}

\vspace {1truecm}
  
\author{G. Gy\"orgyi} 
\email{gyorgyi@glu.elte.hu} 
\affiliation{Institute for Theoretical Physics, E\"otv\"os University,
  1117 Budapest, P\'azm\'any s\'et\'any 1/a, Hungary} 
 
\author{P. C. W. Holdsworth} 
\email{pcwh@ens-lyon.fr} 
\affiliation{ Laboratoire de Physique, Ecole Normale Sup\'erieure,
46 All\'ee d'Italie, F-69364 Lyon cedex 07, France} 
 
\author{B. Portelli}
\email{Baptiste.Portelli@ens-lyon.fr}
\affiliation{ Laboratoire de Physique, Ecole Normale Sup\'erieure,
46 All\'ee d'Italie, F-69364 Lyon cedex 07, France}
  
\author{Z. R\'acz}  \email{racz@poe.elte.hu} \affiliation{Institute for
  Theoretical Physics, E\"otv\"os University, 1117 Budapest,
  P\'azm\'any s\'et\'any 1/a, Hungary} \affiliation{ Laboratoire de
  Physique, Ecole Normale Sup\'erieure, 46 All\'ee d'Italie, F-69364
  Lyon cedex 07, France}
 
\date{July 25, 2003} 

\begin{abstract} 
  
  The extremal Fourier intensities are studied for stationary
  Edwards-Wilkinson-type, Gaussian, interfaces with power-law dispersion.  We
  calculate the probability distribution of the maximal intensity and find
  that, generically, it does not coincide with the distribution of the
  integrated power spectrum (i.\ e.\ roughness of the surface), nor does it
  obey any of the known extreme statistics limit distributions.  The
  Fisher-Tippett-Gumbel limit distribution is, however, recovered in three
  cases: (i) in the non-dispersive (white noise) limit, (ii) for high
  dimensions, and (iii) when only short-wavelength modes are kept. In the last
  two cases the limit distribution emerges in novel scenarios.

\end{abstract}
 
 
\maketitle
 
 

\vspace{1cm}

\section{Introduction}

Extreme value statistics (EVS) has traditionally found applications in
the analysis of environmental and engineering data, such as water
level fluctuations or appliance lifetime~\cite{Coles}.  Recently there
has been a surge of activity from physicists in modeling phenomena
that are governed by the extremal values of some random variable.
Examples range from the low-temperature state in spin glasses
\cite{Bouchaud}, a variety of disordered systems with traveling fronts
(see \cite{Majumdar} for a brief review), and modeling corrosive
fracture \cite{Baldassari}, to depinning of surfaces
\cite{Dahlstedt,Skoe}, to relaxation in granular materials
\cite{Anna}. 
 
A central notion in EVS is the existence of limit distributions, that
is, the extremal value in a batch of a large number of independent,
identically distributed (i.\ i.\ d.) random scalar variables obeys a
probability density function (PDF), which is limited to one of three
main types~\cite{Galambos,Coles, NoteEVS}.  It is the tail of the PDF
of the original variable, the parent PDF, that determines which one of
these three classes the EVS will belong to.  One distinguishes the
Fisher-Tippett-Gumbel (FTG) distribution, for a decay faster than any
power law, the Fisher-Tippett-Fr\'echet (FTF), for power tail, and
Weibull, for power law at a finite edge.  The limit PDFs can even be
cast into a single function with a parameter whose different ranges
correspond to different traditional classes, see e.g.\ 
Ref.~\cite{Coles}.
 
The appearance of EVS limit distributions in physical systems is in
itself interesting and yields a new tool for the description of those
systems, see e.g.\ \cite{Bouchaud}.  On the other hand, there are
instances when the quantities in question are either dependent,
differently distributed, or, have several components and thus
generically do not belong to any of the aforementioned EVS classes,
see e.g.\ \cite{Dahlstedt,Dean,Praehofer}.  Closely related to our
present subject are the studies on the distribution of local
\cite{Toro,Lam,Toro2} and global \cite{RCPS} extrema of the height of
surfaces.  Whereas there is no general mathematical theory for the EVS
of non-i.~i.~d.\ quantities, quite a few special cases have been
clarified, Refs.\ \cite{Galambos,Coles} has some examples of them.
 
A possible connection of EVS to non-Gaussian fluctuations of spatially
averaged, or, global, quantities in critical (strongly correlated) systems has
recently been raised~\cite{BHP98,Bram01,ES,Dahlstedt,ES2,bra.02a,1f} in
relation to both experiments~\cite{Pinton1,BHP98,Pinton2,1f,Lath,Carr} and
simulations~\cite{BHP2000,aum.01,nul.02} for many-body systems as well as for
environmental data~\cite{jan.99,Duna}.  There are two striking characteristics
of this ensemble of observations.  Firstly, the form of the PDF is similar for
these observables~\cite{BHP98,BHP2000}, and secondly, these functions bare a
strong resemblance to the FTG distribution.  There is some variation in form
and there are some clear exceptions, but this approximate universality has
been the subject of considerable debate in the literature.
  
More concrete connections to EVS have been found in studies of the
fluctuations of the roughness, or, width, of interfaces.  The
non-Gaussian nature of these fluctuations was first discussed
in~\cite{FORWZ}, and in~\cite{RP94,PRZ94,Bram01,1f,falpha,Duna}
families of PDFs were identified, all characterized by a single
maximum, positive skewness, an exponential, or near exponential, tail
for large fluctuations above the mean, and even faster decay below it.
In Gaussian interface models relation to the FTG distribution has been
observed in several instances.  Namely, the magnetization distribution
of the $2d$ XY model, in the low-temperature phase where vortices are
rare, has been related to the roughness distribution in the stationary
$2d$ Edwards-Wilkinson interface, and strong resemblance has been
found to a formal generalization of the FTG function~\cite{BHP2000}.
In another work on the $2d$ XY model, the FTG function was numerically
observed for extremal mode selection~\cite{PH}.  A strong connection
to EVS comes from an analytic result on a $1d$ interface model with
long-range interactions, corresponding to a Gaussian noise with $1/f$
power spectrum, where the roughness of the interface turns out to be
exactly of the FTG form~\cite{1f}. 
 
Given the manifold occurrence of the FTG shape, we are led to study
the relation between EVS and the roughness of Gaussian interface
models.  These models can be considered as generalizations of the
stationary $d$-dimensional Edwards-Wilkinson interfaces, specified by
the exponent $\alpha$ in the power-law dispersion of Fourier modes
(for $\alpha=2$ the Edwards-Wilkinson case is recovered).  Due to
their integrability, they are ideal starting point for the study of
more complex, strongly fluctuating systems.  As the models diagonalize
into statistically independent modes, and the roughness equals the sum
of the Fourier intensities, it is straightforward to address the
problem of possible connection to EVS.  Namely, one can pose the
question that originally motivated our research: do the modes with the
largest intensities dominate the PDF of the roughness?  Here we show
that the answer to this question is no.  In fact, this negative
result, the first main conclusion of this paper, is immediately
apparent once we construct the distribution of the maximal intensity,
an expression that turns out to be a generalized form of the Dedekind
function.  Besides being different from the distribution of roughness,
this function generically also deviates from the FTG or any known EVS
limit distributions.  The reason for the deviation from conventional
EVS is that the independent modes have gapless dispersion, and are
strongly non-identically distributed even for higher wavenumbers.  As
a result, for each realization of the interface, the largest intensity
comes from one of only a few soft modes.  The EVS is therefore
effectively coming from a finite-size system, even in the
thermodynamic limit.
 
The main body of the paper concerns the study of the maximal intensity
PDF, which is found to depend on $d$ and $\alpha$.  While there are no
finite critical $d_c$ or $\alpha_c$, marking thresholds to the known
EVS limit distributions, we discover three limits where FTG statistics
sets in: (i) $\alpha\to 0$, (ii) $d\to\infty$, (iii) when only hard
modes with wave vectors beyond a diverging radius $R$ are kept.
Common in these cases is that the thermodynamic limit can be taken
beforehand, $N\to\infty$, where $N$ is the total number of modes, so
one chooses the maximal intensity out of an infinite set, in contrast
to the conventional EVS procedure.  The extremal value PDF becomes
degenerate in all three cases in the sense that the standard deviation
shrinks to zero on the scale of the mean.  This is similar to the
traditional FTG scenario when such degeneracy appears for 
$N\to\infty$.  Now, however, in each case special scaling should be
applied to resolve degeneracy and reveal the FTG distribution.  It is
suggested that the FTG limit in case (iii) is responsible for the
numerically observed fit to the FTG function in Ref.~\cite{PH}. 
  
Finally, we turn to the question how the EVS changes by considering a
different choice of expansion functions (modes). This study further
illuminates the fact that the EVS is not generically described by any
of the traditional EVS limit functions.  Since the dominant
contribution to the EVS comes from a few modes, the EVS will depend on
the specific expansion functions.  We find analytically a different
family of extreme value PDFs for the intensities if the interface is
expanded in sines and cosines, albeit the overall shape goes close to
that of the PDF of the maximal Fourier intensities.
 
We section the paper as follows.  In Section \ref{sec:d1} we define
the Gaussian model (\ref{sec:gauss}), present the basic formulas for
EVS for maximal Fourier intensities (\ref{sec:evs}), then in
\ref{sec:1f} evaluate the EVS for the model showing $1/f$ noise.
General $1/f^\alpha$ noise is considered in \ref{sec:alpha}.  The case
of arbitrary substrate dimensionality is treated in
Sec.~\ref{sec:gendim}, with the general formulas for the extremal
distributions calculated in \ref{sec:gend-evs}.  In Sec.~\ref{sec:XY}
the EVS for the magnetization modes for the XY system in the frame of
the Gaussian model is investigated. The emergence of the FTG
distribution in three limiting cases is described in \ref{sec:univ}.
The EVS of square amplitudes from the expansion in sines and cosines is
discussed in \ref{sec:AFE}, Sec.~\ref{sec:roughness} is devoted to a
comparison to the distribution of the roughness, and Sec.\ 
\ref{sec:concl} contains the concluding remarks.  We give the list of
abbreviations in Appendix \ref{sec:Abbr}, the small-$z$ asymptotes of
the extremal value PDFs are derived in Apps.~\ref{sec:A} and
\ref{sec:A2}, and the finite-$N$ correction to the extremal
distribution is considered in App.~\ref{sec:FS}.  Some details for the
calculation of the PDF in the white noise limit is clarified in
App.~\ref{sec:C}, and in App.~\ref{sec:E} we determine, for
$\alpha>d$, the initial asymptote of the PDF of the roughness for
comparison.
 
 
\section{Gaussian surfaces in one dimension}
\label{sec:d1} 
 
\subsection{$1/f^\alpha$ noise}
\label{sec:gauss}
 
The interface at position $x$ on a one-dimensional ($d=1$) substrate
is characterized by the height $h(x)$.  Replacing $x$ by time $t$,
$h(t)$ can be thought of as the distance from the origin of a random
walker. 
 
The roughness, or mean-square width, is given by
\begin{equation}
w_2(h)=\overline{[\, h(t)-{\overline{h}}\, ]^2}, \label{w2}
\end{equation}
where over-bar denotes the average
of walks, or interfaces on the interval
$0\le t\le T$
\begin{equation}
\overline F =\frac{1}{T}\int_0^T F(t) \td t.
\end{equation} 
The Fourier decomposition of the interface gives the amplitudes of the
fluctuating modes we are interested in 
\begin{equation}
h(t) = \sum_{n=-N}^{N} c_{n} e^{2 \pi i n t/T}\,,\hspace{5mm}
c_{-n} = c_{n}^{\star}.
\end{equation}
Here $h(t)$ is defined on $N_0$ equidistant points ($t=k\Delta t$,
$T=N_0\Delta t$) and we introduced the notation $N=(N_0-1)/2$, with
$N_0$ assumed to be odd, before finally taking the thermodynamic limit
$N\to \infty$.  The roughness is then the integrated power spectrum
\begin{equation} 
w_2(h)= \frac{1}{N}\sum_{n=1}^{N} \left|c_{n}\right|^2.
\label{w2ampl} 
\end{equation}
     
In the models we consider, the time signal $h(t)$ has periodic boundary
condition, it exhibits $1/f^\alpha$ power spectrum, and the modes
are independent Gaussian variables.  The path probability is 
given by 
\begin{equation} 
{\cal P}[h(t)]\propto \exp[{-S[h(t)]}],
\label{probsurf} 
\end{equation} 
where the action $S$ is given in terms of Fourier intensities $\vert
c_n\vert^2$ as
\begin{equation} 
S= \sigma_0 T^{1-\alpha} \sum_{n=1}^{N} n^{\alpha} \vert c_n\vert^2,
\label{effham} 
\end{equation} 
with exponent $\alpha$ defining the $1/f^{\alpha}$ noise spectrum. The
$\sigma_0$ is a parameter setting the effective surface tension and
the power of $T$ originates from dimensional
considerations~\cite{falpha}. 
 
The PDF of the roughness for the Gaussian model has been extensively
studied in \cite{1f,bra.02a,falpha}. For $\alpha=1$ it was found
analytically to coincide with the FTG function \cite{1f}, one of the
limit functions of EVS.  Nevertheless, the roughness is not {\em a
 priori} an extremal quantity and it remains an open question why it
obeys the EVS.  One may argue that the sum is dominated by the softer
modes whose extremal values possibly determine the cumulative behavior
\cite{ES,PH}.  Motivated by this problem we study the largest
contribution to the sum (\ref{w2ampl}). 
  
An alternative choice of Fourier coefficients would be the set
$\left({\cal R}{\rm e} c_{n}\right)^2, \left({\cal I}m
  c_{n}\right)^2$, as studied in~\cite{PH}.  These are the
coefficients from the expansions by sines and cosines, and while this
does not seem to be a physically significant modification, the EVS for
this set is, in general, quantitatively different from that of the
intensities $\left|c_{n}\right|^2$.  Nevertheless, the PDFs for the
two sets are similar in shape and lead to the same physical
conclusions, so in order to keep unity of discussion we focus on the
set $\left|c_n\right|^2$ along most of the main text, and summarize
the results on the $\left({\cal R}{\rm e} c_{n}\right)^2, \left({\cal
    I}m c_{n}\right)^2$ at the end.
 
\subsection{Extreme value distributions}   
\label{sec:evs} 
 
We calculate the probability that the maximal intensity $\vert
c_n\vert^2_{\mathrm{max}}$ is $T^{\alpha-1}z/\sigma_0$, so we shall
actually determine the scaling function of the EVS.  We denote the PDF
for that extreme value by $P_\alpha(z)$, and the cumulative, or,
integrated probability distribution function (IPDF) by
\begin{equation} 
M_\alpha(z)=\int_0^zP_\alpha(y)\td y. 
\label{eq:M} 
\end{equation} 
Since $M_\alpha(z)$ is the probability that none of the
$\left|c_{n}\right|^2$s exceed $T^{\alpha-1}z/\sigma_0$, we can
express it as
\begin{equation}
M_\alpha(z)=\prod_{n=1}^N\int_0^{\left|c_{n}\right|^2\le z}
A_ne^{- n^{\alpha} \vert c_n\vert^2}
\, \td{\cal R}{\rm e}c_n\, \td{\cal I}{\rm m} \,c_n,
\label{ipdf}
\end{equation}
where $A_n=n^{\alpha}/\pi$ is the normalization constant for the PDF
of the $n$th mode.  After evaluating the integrals in (\ref{ipdf})
we find
\begin{equation}
M_\alpha(z)=\prod_{n=1}^N
\left(1-e^{- n^{\alpha} z}\right)\,.
\label{ipdf1}
\end{equation}
In fact, for $\alpha=1$ the product (\ref{ipdf1}) is known from the
defining formula of Dedekind's eta function, the latter also
containing an extra power prefactor~\cite{Apostol}. Despite the
difference in the prefactor, we will refer to (\ref{ipdf1}) as a
generalized Dedekind function.  Differentiating (\ref{ipdf1}) we
arrive at the expression for the PDF of the maximal intensity
\begin{equation} 
P_\alpha(z) =M_\alpha(z) \sum_{n=1}^N  \frac{n^\alpha}{e^{n^\alpha z}-1}. 
\label{pdf1} 
\end{equation} 
The IPDF and the PDF vanish for $z\le 0$, the above formulas 
(\ref{ipdf1},\ref{pdf1}) are understood for non-negative $z$. 
 
Three limits are easily understood here.  Firstly, if $\alpha=0$, i.\ 
e.\ the modes are identically distributed, we find
\begin{equation} 
M_0(z)=\left(1-e^{- z}\right)^N.
\label{Pequi} 
\end{equation} 
Then the change of variables $z=ay+\gamma+\ln{N}$, with
\begin{subequations}\label{eq:ag}
\begin{eqnarray} \label{eq:a}
a&=&\sqrt{\zeta(2)}=\pi/\sqrt{6},\\
\label{eq:gamma}
\gamma&=&\lim_{N\to\infty} \left(\sum_{n=1}^N n^{-1} - \ln N\right),
\end{eqnarray}
\end{subequations}
where $\zeta$ is Riemann's zeta function and $\gamma$ Euler's
constant, yields the FTG distribution in the $N\rightarrow\infty$
limit
\begin{subequations}\label{allGumbel}
\begin{eqnarray}
M_0(z(y)) & \to &  M_{\mathrm{FTG}}(y) = e^{-e^{-ay-\gamma}},
\label{Gumbel-IPDF} \\ 
P_0(z(y))\, z^\prime(y) & \to &  P_{\mathrm{FTG}}(y) =
ae^{-ay-\gamma-e^{-ay-\gamma}}. 
\label{Gumbel}
\end{eqnarray}
\end{subequations}
The constants in (\ref{eq:ag}) were used to scale the FTG distribution
so that the mean becomes zero and the variance one~\cite{NoteFTG}.
Secondly and thirdly, in both limits $\alpha\rightarrow\infty$ and
$z\to\infty$ only the mode $n=1$ matters and one finds
\begin{subequations}
\begin{eqnarray}
M_\alpha(z\to\infty) & \approx & M_\infty(z)  =  1-e^{-z},
\label{exponential-IPDF} \\ 
P_\alpha(z\to\infty) & \approx &  P_\infty(z)  =  e^{-z}.
\label{exponential} 
\end{eqnarray} 
\end{subequations} 
 
Already at this point we can draw one of the main conclusions of the
present paper, namely, that for general $\alpha$ the extreme value
PDFs are none of the known EVS limit functions.  This immediately
follows from the fact that the negative-$z$ and large-$z$ behavior of
(\ref{ipdf1},\ref{pdf1}) is incompatible with those of the limit
functions for the statistics of maxima~\cite{Galambos,Coles}.

The breakdown of validity of the traditional EVS limit distributions
can be ascribed to the fact that, for $\alpha>0$, due to the
$n^\alpha$ dispersion, the distributions of the individual modes are
sufficiently different.  Here ``sufficiently'' needs to be emphasized,
because one can conceive sets of different parent PDFs for the
modes~\cite{parent} that lead to say the FTG function.  Such is the
case of a dispersion that has a gap at the origin or goes to a
constant for high frequencies.  Conversely, one should not be
surprised by the appearance of special PDFs different from the known
limit distributions, because actually for any given PDF one can choose
sets of different parent distributions yielding that PDF in
EVS~\cite{Galambos}. 
  
It is easy to convince oneself that for $\alpha>0$ in formulas
(\ref{ipdf1},\ref{pdf1}) the limit $N\rightarrow\infty$ can be taken
and the resulting distribution has finite moments.  Indeed, because of
dispersion the exponential PDFs of individual modes decay increasingly
fast and so no singularities appear for large $N$.  In conventional
EVS, limit distributions arise because the moments of variable $z$
scale in singular ways in the thermodynamic limit, thus removing most
of the details of the distribution of the original random variables.
Now, however, there is no such singular scale and the strong
dependence on the statistics of the individual modes through the
parameter $\alpha$ remains.  This gives us a physical intuition about
why the EVS for intensities of interfaces with dispersion is not
described by any of the traditional limit distributions.
 
Furthermore, we can also immediately assert the difference between the
PDF of the roughness (\ref{w2ampl}), studied in Ref.~\cite{falpha},
and the PDF (\ref{pdf1}) of the maximal intensity component in
(\ref{w2ampl}).  An obvious deviation is in the physical scales,
namely, while the maximal intensity is always of order $T^{1-\alpha}$,
the mean roughness is like that only for $\alpha>1$, diverges
logarithmically for $\alpha=1$, and stays finite for finite sampling
times $\Delta t$ if $\alpha<1$.  Even when the scaling is the same,
for finite $\alpha>1$, the two PDFs clearly differ (the generating
function of the roughness PDF is given in~\cite{falpha}, whose Laplace
transform is not Eq.~(\ref{pdf1})), and became the same only in the
limit $\alpha\to\infty$, when the $n=1$ mode dominates.  More detailed
comparison will follow in Sec.~\ref{sec:roughness}.

\subsection{$1/f$ noise ($\alpha=1$)}
\label{sec:1f}

\begin{figure}[htb] 
  \psfrag{z/<z>}[c][B][1.25][0]{$z/\!\left<z\right>$}
  \psfrag{<z> P(z)}[B][B][1][0]{$\left<z\right>\,P_1(z)$}
  \psfrag{N=3}[B][B][1][0]{$N=3$\hspace{17pt}~}
  \psfrag{N=6}[B][B][1][0]{$6$} 
  \psfrag{N=9}[B][B][1][0]{$9$} 
  \psfrag{N=12}[B][B][1][0]{$12$} 
  \psfrag{N=50}[B][B][1][0]{$50$} 
  \psfrag{alpha=1, d=1}[B][B][1][0]{} 
  \includegraphics[trim= -50 0 -20 -350, width=6.5cm, angle=-90]{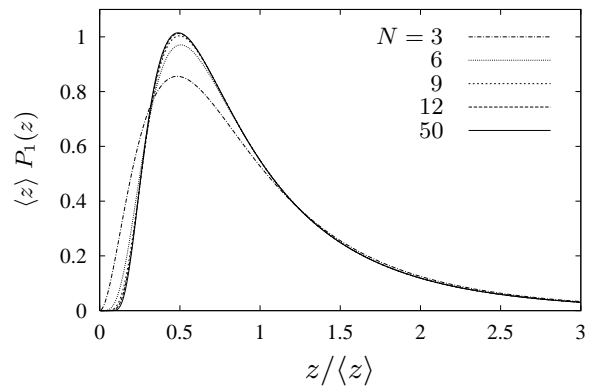} 
  \caption{Distribution functions for the $\alpha=1$ case evaluated numerically
    according to Eq.~(\ref{pdf1}) for various $N$s, each function
    scaled to unit average.  The convergence is apparently fast for
    practical purposes.}
\label{Fig:alpha=1}
\end{figure}

\begin{figure}[htb] 
  \psfrag{z/<z>}[c][B][1.25][0]{$z/\!\left<z\right>$}
  \psfrag{<z> P(z)}[B][B][1][0]{$\left<z\right>\,P_1(z)$}
  \psfrag{N=3}[B][B][1][0]{$N=3$\hspace{17pt}~}
  \psfrag{N=6}[B][B][1][0]{$6$} 
  \psfrag{N=9}[B][B][1][0]{$9$} 
  \psfrag{N=12}[B][B][1][0]{$12$} 
  \psfrag{N=50}[B][B][1][0]{$50$} 
  \psfrag{alpha=1, d=1}[B][B][1][0]{} 
  \includegraphics[trim= -50 0 -20 -350, width=6.5cm, angle=-90]{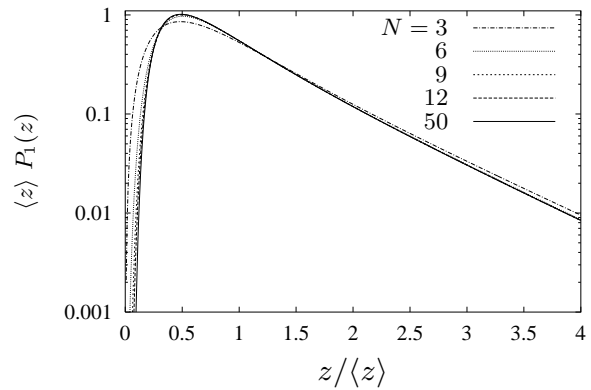} 
  \caption{The same as Fig.~\ref{Fig:alpha=1} on a semilog scale so as
  to better display the regions where $P_1(z)$ is small.}
\label{Fig:alpha=1-log}
\end{figure}

While for large $z$ the PDF $P_1(z)$ is pure exponential as given in
(\ref{exponential}), for general arguments one should resort to
numerical evaluation.  The function (\ref{pdf1}) is shown for a
sequence of $N$s in Figs.~\ref{Fig:alpha=1} and \ref{Fig:alpha=1-log}.
For $N\ge 9$ the curves are hard to distinguish by the naked eye,
the approximation by finite $N$ thus converges fast for practical
purposes.

\begin{figure}[htb] 
  \psfrag{z}[B][c][1.25][0]{$z$}
  \psfrag{P(z)}[B][B][1][0]{$P_1(z)$}
  \psfrag{exact}[B][B][1][0]{exact}
  \psfrag{asymptote}[B][B][1][0]{asymptote\hspace{3pt}~}
  \psfrag{alpha=1, d=1}[B][B][1][0]{}
\includegraphics[trim= -50 0 0 -350, width=6.5cm, angle=-90]{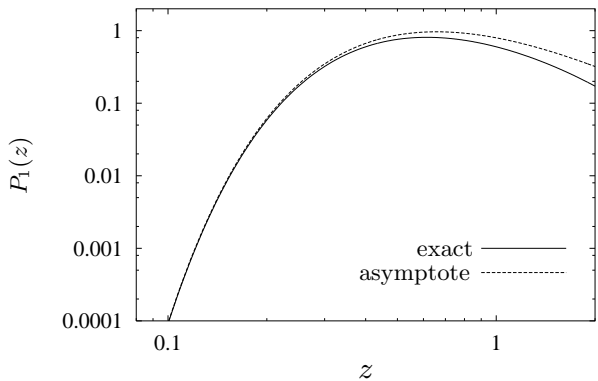}
\caption{The PDF $P_1(z)$ from Eq.~(\ref{pdf1}) with $N=100$ and its
  small-$z$ asymptote as given by Eq.~(\ref{precasymp}).}
\label{Fig:a=1asympt}
\end{figure}

For small-$z$, the expected nonanalytic behavior can be estimated in
the following way: the product in (\ref{pdf1}) can be approximated
as
\begin{equation} 
M_1(z)\approx \prod_{n=1}^{1/z}(nz) 
\approx z^{1/z}\left(\frac{1}{z}\right)!\approx 
\sqrt{\frac{2\pi}{z}}e^{-1/z} 
\label{productapprox} 
\end{equation} 
where we have used Stirling's approximation for the factorial.  While
here we took the terms with $n>1/z$ as one, this gives us a first hint
of the expected functional form.  A more precise calculation in
Appendix \ref{sec:A} shows that the large-$n$ region gives a
contribution of equal order, and in the end the true asymptote differs
from (\ref{productapprox}) only in a scale factor in the exponent
\begin{equation} 
M_1(z)\approx \sqrt{\frac{2\pi}{z}}e^{-\pi^2/6z}, 
\label{productfinal} 
\end{equation} 
see Eq.\ (\ref{eq:A-M-final}) with (\ref{eq:A-c}) for $\alpha=1$.
Derivation by $z$ gives the small-$z$ asymptote of the PDF
\begin{equation} 
P_1(z)\approx \frac{\sqrt{2\pi}\,\pi^2}{6z^{5/2}}e^{-\pi^2/6z}.
\label{precasymp} 
\end{equation} 
As demonstrated in Fig.\ \ref{Fig:a=1asympt}, the above expression is
correct at small $z$ and provides a reasonable approximation over most
of the ascending part of $P_1(z)$. 
 
We conclude the case of the $1/f$ noise by noting that, the PDF for
the roughness (\ref{w2ampl}) scales logarithmically in $N$ and
approaches the FTG function \cite{1f}.  In contrast, as shown above,
direct extremal selection of the constituent intensities in
(\ref{w2ampl}) leads to a nonsingular PDF, related to the Dedekind
function, even in the thermodynamic limit.  In short, we have the FTG
distribution for the roughness, which is not an extremal quantity,
while the EVS is not described by the FTG function. Thus the question
raised in Ref.~\cite{1f}, namely, what kind of extremal value
selection can possibly be responsible for the FTG distribution of the
roughness, has not been answered and is left to further explorations. 
 
\subsection{General $\alpha$} \label{sec:alpha} 
  
Here we go beyond the $1/f$ spectrum and investigate the noise for 
general $\alpha$.  The special cases $\alpha=0$ and $\alpha=2$ 
correspond to white noise and the Wiener process (ordinary random 
walk), respectively.  The PDF of the roughness for Gaussian noise for 
periodic and ``window'' (bulk) boundary condition has been studied in 
\cite{falpha,Duna}, and no $\alpha$ other than $1$ has been found for 
which the PDF coincided with any of the known limit PDFs of extreme 
statistics. 
 
From the previous discussion it is clear that it is only for $\alpha =
\alpha_c=0$ that EVS is given by the FTG function in finite
dimensions.  Concerning the small-$z$ asymptote, the IPDF is given in
Appendix \ref{sec:A} in (\ref{eq:A-M-final}), whence differentiation
yields the PDF
\begin{equation}  
P_\alpha(z)\approx  
\frac{(2\pi)^{\alpha/2}c(\alpha)}{z^{3/2+1/\alpha}} 
\exp\left(-\frac{c(\alpha)}{z^{1/\alpha}}\right). 
\label{alphaasymp} 
\end{equation} 
where $c(\alpha)$ is given in Eq.~(\ref{eq:A-c}).  One can see that
the above form becomes singular only in the $\alpha \rightarrow 0$
limit.  While for any finite $\alpha>0$ the asymptote is obviously
incompatible with the FTG function (\ref{Gumbel}), surprisingly,
expression (\ref{alphaasymp}) corresponds to the generalized FTF
function for $k^{th}$ maximum \cite{Galambos} in the special case when
$k$, the FTF power parameter $\mu$, and $\alpha$ are related through
$\mu=2(k-1)=1/\alpha$.  The function (\ref{alphaasymp}) does not equal
the EVS distribution for larger $z$, it is not even normalized to one,
so this coincidence does not contradict the claim that the extreme
value PDFs $P_\alpha(z)$ are none of the known limit distributions of
EVS.  The FTF class is not expected to be of relevance here anyhow,
because the parent PDFs of the constituent modes decay exponentially. 
  
When one evaluates $P_\alpha(z)$, the convergence in $N$ is important.
The $N$-dependence has been determined in App.~\ref{sec:FS}, whence it
is apparent that convergence is fast for $\alpha$ of unit order and
larger, but slows down for smaller $\alpha$.  We give here the
asymptote in the region of slow decay, $\alpha<1$,
\begin{eqnarray} 
  \label{eq:fs-1d}
  P_\alpha(z)-P_{N,\alpha}(z) \approx M_\alpha(z)\frac{N}{\alpha\,z}
  e^{-N^\alpha z}, 
\end{eqnarray} 
where we kept the $N$ index for $P$ when only $N$ modes were
counted.
 
\begin{figure}[htb] 
  \psfrag{z/<z>}[c][B][1.25][0]{$z/\!\left<z\right>$}
  \psfrag{<z> P(z)}[B][B][1][0]{$\left<z\right>\,P_\alpha(z)$}
  \psfrag{d=1}[B][B][1][0]{}
  \psfrag{a=0.1}[B][B][1][0]{~\hspace{15pt}$\alpha=0.1$}
  \psfrag{0.2x}[B][B][1][0]{$0.2$\hspace{3pt}~}
  \psfrag{0.5x}[B][B][1][0]{$0.5$} 
  \psfrag{1x}[B][B][1][0]{~\hspace{-5pt}$1$} 
  \psfrag{2x}[B][B][1][0]{~\hspace{-5pt}$2$} 
  \psfrag{5x}[B][B][1][0]{~\hspace{-5pt}$5$} 
\includegraphics[trim= -50 0 -20 -350, width=7cm, angle=-90]{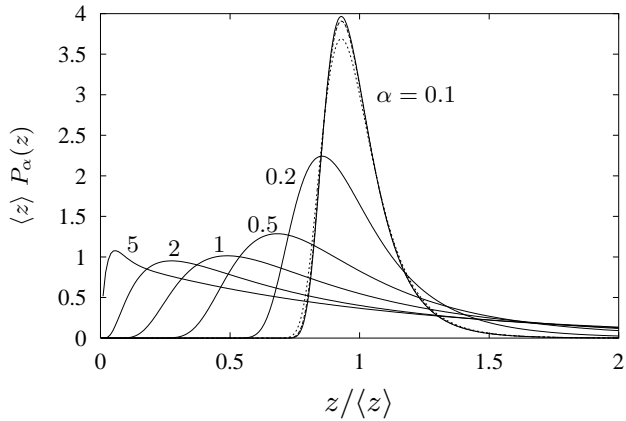}
\caption{Distribution functions for $d=1$ and various $\alpha$s.  We rescaled
  $z$ by its mean $\left<z\right>$ to get a variable of unit average.  For
  $\alpha=0.1$ we show the sequence of approximants with $N=10^5, 10^6, 10^7$,
  peak moving upwards with increasing $N$, to demonstrate slow convergence.}
\label{pdf-d1-a-m}  
\end{figure}

\begin{figure}[htb]  
  \psfrag{z/<z>}[c][B][1.25][0]{$(z-\left<z\right>)/\Delta z$} 
  \psfrag{<z> P(z)}[B][B][1][0]{$\Delta z\, P_\alpha(z)$} 
  \psfrag{d=1}[B][B][1][0]{} 
  \psfrag{a=5}[B][B][1][0]{~\hspace{45pt}$\alpha=5$}
  \psfrag{2x}[B][B][1][0]{$2$}
  \psfrag{0.1}[B][B][1][0]{$0.1$}
  \psfrag{FTG}[B][B][1][0]{FTG}
  \includegraphics[trim= -50 0 -20 -350, width=6.8cm, 
  angle=-90]{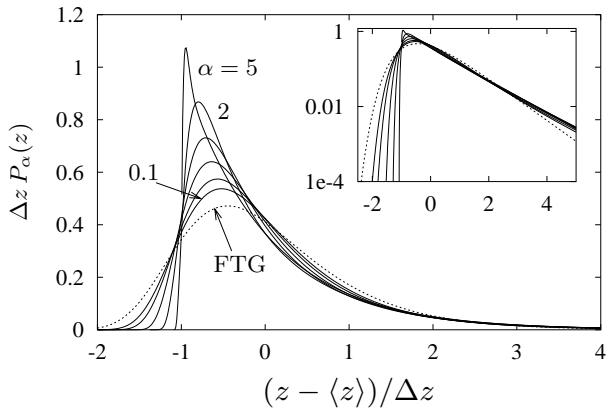} 
\caption{The PDF with the same $\alpha$s as in Fig.~\ref{pdf-d1-a-m}, but $z$
  is rescaled now by its mean $\left<z\right>$ and standard deviation $\Delta
  z$ to produce a variable of zero average and unit deviation.  For
  $\alpha=0.1$ only the curve with the largest $N$ is shown.  The peaks
  decrease with $\alpha$.  The PDFs tend to the FTG function (dashed line). 
  The semilog inset magnifies the small-$P$ regions. }
\label{pdf-d1-a-v}  
\end{figure}

The results of the evaluation of $P_\alpha(z)$ are given in
Figs.~\ref{pdf-d1-a-m} and \ref{pdf-d1-a-v}.  If displayed on the
scale when the mean is set to one, see Fig.~\ref{pdf-d1-a-m}, for
vanishing $\alpha$ the PDFs develop a singularity as eventually they
approach the Dirac delta.  Slow convergence in $N$ for small $\alpha$
is also demonstrated, already for $\alpha=0.1$ one has to go up to
exceedingly large $N$s to get a satisfactory approximation for the
PDF.  When the PDFs are shown with variance scaled to one, as in
Fig.~\ref{pdf-d1-a-v}, the functions remain nonsingular and tend
towards the FTG distribution.

\section{General dimension}
\label{sec:gendim}

\subsection{Extremal intensities}
\label{sec:gend-evs}

It is easy to generalize the above calculations to surfaces $h({\bf
  r})$ defined on a hypercubic lattice substrate of dimension $d$ and
edge length $L$.  We retain the periodic boundary condition for
substrate, so the natural expansion of $h({\bf r})$ involves again
Fourier modes.  The probability of a surface (\ref{probsurf}) is then
characterized by the effective action~\cite{bra.02a,falpha}
\begin{equation} 
S= \sigma_0\,L^{d-\alpha}\,\sum^N_{\nsd}\pr \, |{\nd}|^{\alpha} \vert
c_{\nsd}\vert^2.  
\label{effhamgend}
\end{equation} 
Here the rescaled wave vector ${\nd}=(n_1,n_2,...,n_d)$ has integer
components $n_i$ such that $\left|n_i\right|\le N$, furthermore, the
mark $~^\prime$ implies that if an $\nd$ is counted then $-\nd$ is
not, and the zero vector is excluded.  (The halving of the Brillouin
zone is the consequence of the relation 
$c_{\nsd}=c^{\ast}_{{-\nsd}}$.)  The above action for $\alpha=2$
corresponds to the stationary state of the original Edwards-Wilkinson
model, while for $\alpha=4$ it gives the curvature-driven
Mullins-Herring interface.  For a general $\alpha=2k$, $k$ integer, it
is a Gaussian massless model with finite-range, and for $\alpha\neq
2k$ with interactions decaying like a power law. 
  
Denoting the maximal value of $\sigma_0\,L^{d-\alpha}\,\vert
c_{\nsd}\vert^2$ by $z$, and its IPDF by $M_{\alpha,d}(z)$, each mode
gives a multiplicative factor $1-e^{- |{\nsd}|^{\alpha} z}$ resulting
in
\begin{eqnarray}
  \label{eq:B-M}
  M_{\alpha,d}(z) = \prod^N_{\nsd}\pr \left( 1-e^{-|{\nsd}|^\alpha z}\right),
\end{eqnarray}
whence the PDF is
\begin{equation}
P_{\alpha,d}(z) = M_{\alpha,d}(z)\,{\sum^N_{\nsd}}\pr
\frac{|{\nd}|^\alpha}{e^{|{\nsd}|^\alpha z}-1}. 
\label{pdfd}
\end{equation} 

The IPDF (\ref{eq:B-M}) can be considered as a further generalization
of Dedekind's original product formula~\cite{Apostol}.  As in one
dimension, it is straightforward to show that for $\alpha>0$ the above
functions remain finite and involve finite mean and variance in the
limit $N\to\infty$.  Thus again no singular scaling is necessary and
so we do not expect that any of the known extreme value limit PDFs
emerge for general $d$ and $\alpha$.  However, again as in $d=1$, for
$\alpha=0$ all independent modes become identically distributed and we
recover the FTG function after proper scaling by $N$.

The large $\alpha$ limit of (\ref{pdfd}), for any fixed dimension
$d$, is determined by the contribution of the modes $|{\nd}|=1$
\begin{eqnarray}
P_{\infty,d}(z)  =d\,e^{-z} \left(1 - e^{-z} \right)^{d-1}.
\label{genasympt1}
\end{eqnarray}
For finite $\alpha$ where all modes count, we have closed forms
only for the asymptotes.  In the large-$z$ limit, for any
$\alpha>0$ and fixed $d$, one obtains
\begin{equation}
P_{\alpha,d}(z\rightarrow \infty)  \approx d\,e^{-z}.
\label{genasympt2}
\end{equation}
Since here, too, only the modes $|{\nd}|=1$ matter, the large-$z$
formula is obtained from the PDF (\ref{genasympt1}) for
$\alpha\to\infty$, by taking $z\to\infty$, which explains the fact
that (\ref{genasympt2}) is independent of $\alpha$.  For $z\to 0$ we
determined, in Appendix \ref{sec:B1} the asymptote of the logarithm of
the IPDF $M_{\alpha,d}(z)$, see Eqs.\ (\ref{eq:B-logM}, \ref{eq:B-c}).
One can easily convince oneself that the leading term for the
logarithm of $P_{\alpha,d}(z)$ is the same as for $M_{\alpha,d}(z)$,
so we have for small $z$
\begin{eqnarray}  
  \label{eq:logP} 
\ln P_{\alpha,d}(z) \approx -   \frac{\pi^{d/2}
 \zeta\left(1+\frac{d}{\alpha}\right) \, 
 \Gamma\left(1+\frac{d}{\alpha}\right)}{z^{d/\alpha}\, d\, 
 \Gamma\left(\frac{d}{2}\right)}.
\end{eqnarray} 
For general $z$ we have to evaluate expression (\ref{pdfd})
numerically. This poses no difficulties unless $\alpha\approx 0$.

\subsection{The $2d$ Edwards-Wilkinson model ($\alpha=2$)}
\label{sec:XY}

\begin{figure}[htb]
  \psfrag{z/<z>}[c][B][1.25][0]{$z/\!\left<z\right>$}
  \psfrag{<z> P(z)}[B][B][1][0]{$\left<z\right>\,P_{2,2}(z)$}
  \psfrag{alpha=2, d=2}[B][B][1][0]{}
  \psfrag{PDF}[B][B][1][0]{$P_{2,2}$}
  \psfrag{FTG}[B][B][1][0]{FTG}
\includegraphics[trim= -50 0 -20 -350, width=6.8cm, angle=-90]{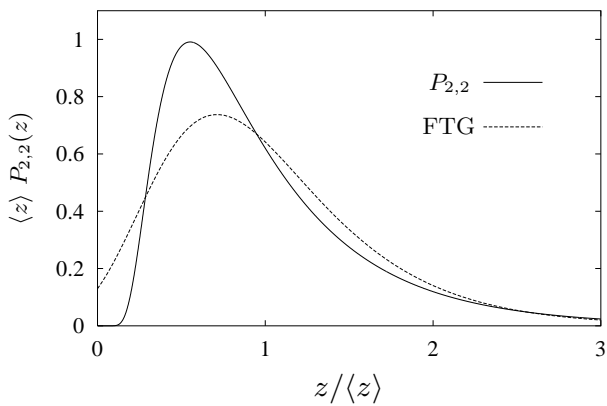}
\caption{Distribution function for  the maximum
  Fourier intensity statistics for $d=\alpha=2$ and the FTG function
  scaled to unit mean and the same variance.  }
\label{Fig:compXY}
\end{figure}

One of the physically most relevant cases is $d=2$ and $\alpha=2$, the
Edwards-Wilkinson surface as originally defined.  Since it is a
massless Gaussian system, it also corresponds to the XY model in that
part of the low temperature phase where the effect of vortices is
negligible~\cite{ES}.  There the magnetization distribution is
proportional, with a change of sign, to the roughness of the
Edwards-Wilkinson surface.  Accordingly, the distribution of the
maximum amplitude of the magnetization fluctuations, after proper
scaling, should be given by the PDF (\ref{pdfd}) for $d=\alpha=2$.
Fig.~\ref{Fig:compXY} shows the $P_{2,2}(z)$, together with the FTG
function normalized to the same mean and variance.  While the fact
that the two functions are different is obvious, here we demonstrate
that the functions deviate significantly and in no range could they be
mistaken for each other.  The large difference here is of importance,
because the PDF $P_{2,2}(z)$ characterizes also the XY model, so this
is an example when the FTG statistics is far from an EVS in a critical
many-body system.

\section{FTG limiting  cases}
\label{sec:univ} 

\subsection{Case (i): White noise limit $\alpha\to 0$}
\label{sec:alpha0} 

As we have seen, for $\alpha=0$ the modes become identically
distributed in Eq.\ (\ref{eq:B-M}), thus, because of the exponential
decay of the parent PDF for large $z$, one recovers the FTG function
by proper scaling in $N$.  One may, therefore, expect that when
$N=\infty$ is set first, the extreme value distribution converges to
FTG for $\alpha\to 0$, if proper scaling in $\alpha$ is applied.  Such
a convergence is not part of the standard theory of the FTG limit
\cite{Coles,Galambos}, so we show here how it comes about.  A
numerical demonstration in $d=1$ was given before on Fig.\ 
\ref{pdf-d1-a-v}, now we derive it analytically for any $d$, and
determine the natural scaling of the variable $z$ by $\alpha$.

Let us start out from the IPDF given by (\ref{eq:B-M}).
Exponentiating the product into a sum, we realize that in the
interesting region of $z$, where $-\ln M(z)$ is not too large, $z$
must be large, thus we can linearize the logarithms.  Then we notice
that for small $\alpha$ the terms in the sum change slowly with
$|{\nd}|$, therefore we assume that the sum can be replaced by an
integral.  So we get
\begin{eqnarray}
\ln M_{\alpha\to 0,d}(z) &\approx& - \frac{1}{2} \sum_{|{\nsd}|>0}
e^{-|{\nsd}|^\alpha z} \nonumber \\ &\approx& -
\frac{\pi^{d/2}}{\Gamma(d/2)} 
\int_1^\infty e^{-n^\alpha z}\, n^{d-1}\td n.
\label{eq:a0M1}
\end{eqnarray}
Asymptotic analysis of this expression is done in Appendix
\ref{sec:C}, and results in
\begin{eqnarray}
\ln M_{\alpha\to 0,d}(z) &\approx& - A
\left(\frac{\xi}{z}\right)^{d/\alpha},
\label{alpha0lnM}
\end{eqnarray}
where 
\begin{subequations}
\label{eq:a0}
\begin{eqnarray}
\xi &=& \frac{d}{\alpha\,e},
\label{alpha0xi}  \\
A &=& \frac{ \pi^{d/2}}{\Gamma(d/2)}\, \sqrt{\frac{2\pi}{\alpha\,d}},
\label{alpha0A}
\end{eqnarray}
\end{subequations}
valid in the region where $z$ is $\xi$ in leading order.  Introducing
the variable $y$ by the linear transformation (for the constants
$a,\gamma$ see Eq.~(\ref{eq:ag}))
\begin{equation}
  \label{eq:a0z}
  z = \xi + \frac{1}{e}\left(ay+\gamma+\ln A\right),
\end{equation}
and keeping $y$ of order unity when $\alpha\to 0$ we get
\begin{equation} 
  \label{eq:a01}
A\,\left(\frac{\xi}{z}\right)^{d/\alpha} \approx e^{-ay-\gamma},
\end{equation} 
up to terms vanishing with $\alpha$.  Since (\ref{alpha0lnM}) was the
leading term in the logarithm of the IPDF we get
\begin{equation} 
  \label{eq:a0M}
   M_{\alpha\to 0,d}(z) \approx e^{-e^{-ay-\gamma}},
\end{equation}
which is the IPDF (\ref{Gumbel-IPDF}) for the FTG-distributed variable
$y$.  Since this IPDF has zero mean and unit variance, the linear
transformation (\ref{eq:a0z}) is equivalently
\begin{eqnarray} 
  \label{eq:zy} 
  z = \left< z \right> + y \Delta z, 
\end{eqnarray} 
where $\left< z \right>$ is the mean and $\Delta z$ the standard
deviation of $z$ up to terms vanishing for $\alpha\to 0$.  It thus
follows that the scaled maximal intensities $z$ have an average
diverging proportionally to $\alpha^{-1}$, and they scatter in an
${\cal O}(1)$ region about the average. 
 
\subsection{Case (ii): High dimensions} 
\label{sec:hd}  
 
We immediately recognize the FTG limit when $d\to\infty$ in the
formula (\ref{genasympt1}) of the PDF for large $\alpha$,
$P_{\infty,d}(z)$.  Indeed, the PDF for the variable $y=(z-\gamma-\log
d)/a$ then goes over to the FTG limit function (\ref{Gumbel}).  Below
we study the $d\to\infty$ limit for any fixed positive $\alpha$ and
conclude that again FTG arises.  This is far from obvious {\em a
  priori}, because while in large dimensions there are many modes in
each shell of constant $|{\nd}|$, so they have the same, exponential,
parent distribution and, if we had only them, they would give rise to
FTG, but on different shells the intensities are non-i.\ i.\ d.\ 
variables.
 
Again we exponentiate the product (\ref{eq:B-M}) into a sum, and
notice that if $d$ is large, the entropic weight quickly increases
with $|{\nd}|$.  Thus the dominant contribution comes from large
$|{\nd}|$s, so we replace the sum by an integral and again get formula
(\ref{eq:a0M1}), which should be taken now for any fixed $\alpha>0$
but in the limit $d\to\infty$.  Its asymptotic analysis is similar to
that described for $\alpha\to 0$ in Appendix \ref{sec:C}, and gives
\begin{eqnarray}
\ln M_{\alpha,d\to\infty}(z) &=& - \frac{1}{\sqrt{2\alpha}}\, 
\left(\frac{\xi}{z}\right)^{d/\alpha}, \label{highdlnM}
\end{eqnarray} 
where
\begin{eqnarray}
\xi &=& \frac{(2\pi e)^{\alpha/2}}{\alpha\,e}
\, d^{1-\alpha/2},  
\label{highdxi} 
\end{eqnarray} 
provided $z$ is of the order of $\xi$.  Next we introduce $y$ through
\begin{equation} 
  \label{eq:highdz} 
  z = \xi \left[ 1+ \frac{\alpha}{d}\left(ay+\gamma-\frac{\ln
  2\alpha}{2}\right) \right].
\end{equation} 
Keeping $y$ at order unity while $d\to\infty$, we get
\begin{equation} 
  \label{eq:highd1}
\frac{1}{\sqrt{2\alpha}}\,
\left(\frac{\xi}{z}\right)^{d/\alpha} \approx e^{-ay-\gamma}, 
\end{equation} 
whence (\ref{highdlnM}) yields the IPDF
\begin{equation}  
  \label{eq:highdM}
    M_{\alpha,d\to\infty}(z) = e^{-e^{-ay-\gamma}},
\end{equation} 
which coincides with the FTG distribution (\ref{Gumbel-IPDF}).
 
We can thus conclude that in high dimensions the extremal intensities
belong to the FTG class.  Furthermore, by comparing Eq.~(\ref{eq:zy})
with (\ref{eq:highdz}) we can determine the average and standard
deviation of the scaled maximal intensity $z$, namely,
$\left<z\right>\propto d^{1-\alpha/2}$ and $\Delta z\propto
d^{-\alpha/2}$.  Interestingly, $\left<z\right>$ diverges only for
$\alpha<2$, while it shrinks to zero when $\alpha>2$, so the latter
case is an example of a nonconventional FTG limit, when the maximal
intensities are small although the intensities themselves are not
bounded from above.  While it is plausible that strong enough
dispersion can have the effect of an upper cutoff on the amplitude,
the novelty here is the sharp transition at $\alpha=2$, the only value
when the characteristic maximal intensities remain finite and
positive.  For all $\alpha>0$, however, the scale of the standard
deviation becomes much smaller than that of the average, a feature of
the conventional FTG scenario~\cite{Coles,Galambos}.
 
\subsection{Case (iii): Hard modes} 
\label{sec:hm} 
 
We study the situation when the maximal amplitude is selected only
from among those with $|{\nd}|\ge R\gg 1$, that is, the very hard
modes.  We consider arbitrary but fixed $\alpha$ and $d$.  For we take
the thermodynamic limit $N\to \infty$, we are left with a single
divergent parameter $R$.  Whereas all hard mode intensities are quite
small, they are of different scales and so are essentially
non-i.i.d. variables.  Thus special considerations are necessary to
determine their EVS.

The IPDF for the maximal hard mode amplitude is 
\begin{eqnarray}  
\label{eq:HM-M} 
  M_{\alpha,d}(z,R) ={\prod_{|{\nsd}| \ge R}}\!\!\!\pr \left(
  1-e^{-|{\nsd}|^\alpha z}\right). 
\end{eqnarray}
Obviously $R=1$ gives the formerly studied IPDF (\ref{eq:B-M}) in the
thermodynamic limit.  The interesting region in $z$ is where $M(z)$
changes fast.  At this stage we assume that in that region $R^\alpha
z$ is large, but we should check the result for consistency in the
end.  We can make from (\ref{eq:HM-M}) a sum in the exponent, and
since large $|{\nd}|$s are involved, we rewrite the sum into an
integral.  In leading order we obtain
\begin{eqnarray} 
  \label{eq:HM-lnM} 
   \ln M_{\alpha,d} (z,R\to\infty) &\approx& - \frac{B\,
    R^{d-\alpha}}{z} \exp\left(-R^\alpha z\right), 
\end{eqnarray} 
where 
\begin{eqnarray} 
    B&=&\frac{\pi^{d/2}}{\alpha\,\Gamma(d/2)}.   \label{eq:HM-B}
\end{eqnarray} 
Careful consideration of the compounding logarithmic
singularities leads to the observation that in terms of the variable
$y$, introduced by
\begin{eqnarray} 
  \label{eq:HM-y} 
    z = R^{-\alpha}\left( ay + \gamma + \ln \frac{B}{d}   + d \ln R -
    \ln \ln R\right),
\end{eqnarray}
the expression (\ref{eq:HM-lnM}) is just $-e^{-ay-\gamma}$ up to terms
vanishing with increasing $R$.  That is, we have recovered the FTG
function (\ref{Gumbel-IPDF})
\begin{equation} 
  \label{eq:HM-FTG}
   M_{\alpha,d}(z,R\to\infty) \approx e^{-e^{-ay-\gamma}},
\end{equation} 
where the linear transformation (\ref{eq:HM-y}) is understood.
Comparing (\ref{eq:HM-y}) with (\ref{eq:zy}) we find that the mean
$\left< z\right>$ has the leading singularity $R^{-\alpha}\ln R$, so
in the relevant region of $z$ the $R^\alpha z$ indeed diverges, as
assumed in the above derivation.

\subsection{Scales of singularity} 
\label{sec:sc} 
 
Summarizing the aforementioned limits, we found that FTG emerges (i)
as $\alpha\to 0$, (ii) for $d\to\infty$, and (iii) when only hard
modes $R\le |{\nd}|$ with $R\to\infty$ are considered.  While for
$\alpha=0$ the intensities become i.\ i.\ d.\ and so FTG should be
expected, they are {\em a priori} non-i.\ i.\ d.\ in the cases
(ii,iii).  Nevertheless, for $d\to\infty$ we found that a shell of
practically i.\ i.\ d. modes in the Brillouin zone becomes dominant,
so this essentially explains why one of the known EVS limit
distributions emerged.  On the other hand, in the case of hard modes
we do not see i.\ i.\ d.\ intensities grouping, thus presently we lack
an intuitive explanation for FTG.  Remarkably, however, a common
feature of all the above cases is that the distribution narrows down
to a scale smaller than that of the average, a phenomenon also present
in the traditional FTG scenario.  In Table \ref{tab:univ} we summarize
the scales of the mean and standard deviation of the maximal
amplitude, and compare them to the scales in the conventional limit
when FTG emerges.  In the latter case we consider batches of i.\ i.\ 
d.\ variables, whose parent IPDF approaches $1$ like $\exp\left(-a\,
  z^r\right)$ for large $z$.  The divergent parameter is then the
number $N$ of variables in a batch.  Then the statistics of the
maximal values within the batches becomes of FTG type \cite{Galambos},
and straightforward calculation yields the scales given in the last
row of the table.
\begin{table}
\caption{\label{tab:univ} Order of the mean and standard deviation of
  the scaled maximal intensity $z$ in the FTG limit cases (i-iii).
  For comparison the scales in the  traditional scenario 
  are also shown, see text for the parameters $N,r$.}   
\begin{ruledtabular}
\begin{tabular}{lccc}
Case & $\left< z\right>$  & $\Delta z = \sqrt{\left< z^2\right> -  \left<
    z\right>^2}$  & $\Delta z/\left< z\right>$ \\ 
\hline
(i) ~~~$\alpha\to 0$ & $\alpha^{-1}$ & $1$ & $\alpha$ \\
\hline
(ii) ~~$d\to\infty$  & $d^{1-\alpha/2}$ & $d^{\,-\alpha/2}$ &  $d^{-1}$\\
\hline
(iii) ~$R\to\infty$  & $R^{-\alpha}\ln R$ & $R^{-\alpha}$  & $(\ln R)^{-1}$\\
\hline
i.i.d. $\!N\to\infty$  & $(\ln N)^{1/r}$ & $(\ln N)^{1/r-1}$ &   $(\ln N)^{-1}$ \\
\end{tabular} 
\end{ruledtabular}  
\end{table}

\section{Sine and cosine expansion} 
\label{sec:AFE} 
 
Based on physical intuition one may suspect that, in the absence of
the traditional EVS limit for i.\ i.\ d.\ variables, the extremal
amplitude PDFs will depend on the choice of expansion functions.
Given the periodic boundary conditions for the surface, another
natural choice of expansion functions are the sines and cosines, whose
coefficients squared are $\left({\cal R}{\rm e} c_{\nsd}\right)^2,
\left({\cal I}m c_{\nsd}\right)^2$.  Below we show that indeed a new
family of PDFs arise for the maximal square amplitude in this case,
thus further illustrating deviation from the known EVS limit
functions.  More motivation to look at these coefficients comes from
the fact that the first EVS study in this area was done with such an
expansion in Ref.~\cite{PH}.  There, the FTG distribution was found
numerically when soft modes were discarded.

When the sine and cosine modes are considered separately, we obtain
the IPDF of $z\, L^{\alpha-d}/\sigma_0$ being the maximum of ${\cal
  R}{\rm e}^2c_{\nsd}$ and ${\cal I}{\rm m} ^2c_{\nsd}$ as
\begin{equation}
{\tilde M}_{\alpha,d}(z)={\prod^N_{\nsd}}\pr\,\int_0^{\sqrt{z}} \int_0^{\sqrt{z}} 
A_ne^{- \left|{\nsd}\right|^{\alpha} \vert c_{\nsd}\vert^2}
\, \td{\cal R}{\rm e}\,c_{\nsd}\, \td{\cal I}{\rm m} \,c_{\nsd}. 
\label{M21}
\end{equation}
 Then one straightforwardly gets
\begin{eqnarray} 
  \label{eq:Mgenerald2} 
  {\tilde M}_{\alpha,d}(z) = {\prod^N_{\nsd}}\pr \, {\rm erf}^2 \!\left(\sqrt{|{\bf
  n}|^\alpha z}\right),
\end{eqnarray} 
whence 
\begin{equation}
{\tilde P}_{\alpha,d} =\frac{2 {\tilde M}_{\alpha,d}(z)}{\sqrt{\pi z}} {\sum^N}\pr\,
\frac{|{\nd}|^{\alpha/2} e^{- \left|{\nsd}\right|^{\alpha} z} }
{{\rm erf}
  \left(\sqrt{|{\nd}|^{\alpha} z}\right)}. 
\label{pdfd2}\end{equation}
This will be the basis for numerical evaluation where we go up to
an $N$ where the curve in the figure visibly stabilizes.

For $\alpha\rightarrow\infty$ the $\left|{\nd}\right|=1$ modes
numbering $2d$ dominate, so
\begin{eqnarray}
{\tilde P}_{\infty,d}(z)  &= &\frac{2\,d\, e^{-z}}{\sqrt{\pi z}}\, {\rm
  erf}^{2d-1}\! \left(\sqrt{z}\right), 
\label{alphainftyg2} 
\end{eqnarray}
The large-$z$ formula for any $\alpha$ is also determined by the
softest modes, so it is just the asymptote of (\ref{alphainftyg2})
\begin{equation}
{\tilde P}_{\alpha,d}(z\to\infty) \approx \frac{2\,d\, e^{-z}}{\sqrt{\pi z}},
\label{zinftyg2} 
\end{equation}
while for $z\to 0$ the logarithm of the PDF has the same leading term as the
IPDF, given in Eqs.\ (\ref{eq:B-logM2}, \ref{eq:B-c2}).  
   
\begin{figure}[htb] 
  \psfrag{z/<z>}[B][B][1.25][0]{$z/\!\left<z\right>$}
  \psfrag{<z> P(z)}[B][B][1][0]{$\left<z\right>\,{\tilde P}_1(z)$}
  \psfrag{alpha=1, d=1}[B][B][1][0]{}
  \psfrag{P11}[B][B][1][0]{${\tilde P}_1$}
  \psfrag{Pt1}[B][B][1][0]{${P}_1$}
\includegraphics[trim= -50 0 0 -350, width=6.5cm, angle=-90]{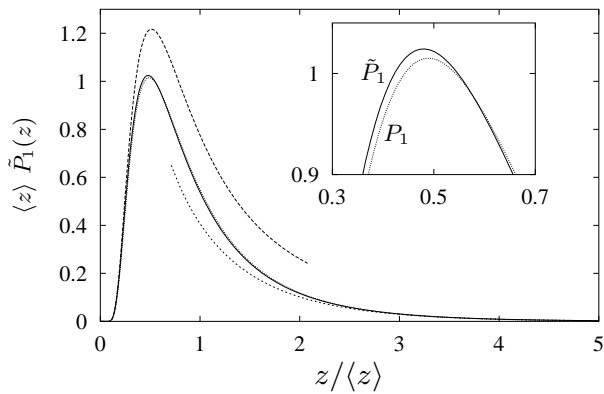} 
\caption{The extreme value PDF ${\tilde P}_{1}(z)$, computed from
  (\ref{pdfd2}) with $\alpha=d=1$ (full line), the asymptotes for small $z$
  from (\ref{zinftyg2}) (dashed), and for large $z$ (\ref{precasymp2}) (short
  dashed).  For comparison the maximal intensity PDF $P_{1}(z)$ from
  Fig.~\ref{Fig:alpha=1} is also shown (dotted); the two PDFs are very close,
  but distinct, as can be observed also near the maximum (see inset).}
\label{Fig:a1d1ii}   
\end{figure}

For the sake of demonstration we consider $\alpha=d=1$, and denote the
corresponding PDF by ${\tilde P}_1$.  For small $z$, Eq.~(\ref{eq:A-M2-final})
of Appendix \ref{sec:A2} with $\alpha=1$ gives the asymptote
\begin{equation} 
{\tilde P}_1(z)\approx \frac{\pi\,c_1}{\sqrt{2}\,z^{5/2}}e^{-c_1/z},
\label{precasymp2}     
\end{equation}    
where $c_1={\tilde c}(1,1)=1.3320405$ was computed from
(\ref{eq:B-c2}).  The full function and the asymptotic formulas are
illustrated in Fig.~\ref{Fig:a1d1ii}.  Note that the deviation of the
small-$z$ asymptote from the exact PDF never exceeds $0.25$ even for
larger $z$s.  We also displayed the PDF of extremal intensity $P_1(z)$
from Fig.~\ref{Fig:alpha=1}, which is given by a different formula and
has a different mean $\left<z\right>$, but after rescaling goes
surprisingly close to $\tilde P_1$.  Nevertheless, the two functions
can still be distinguished as shown on the inset of
Fig.~\ref{Fig:a1d1ii}.  This demonstrates that in the present case the
EVS depends weakly on the expansion functions.
  
Interestingly, in the three limits yielding the FTG distribution for
the intensities $\left| c_i\right|^2$ in Sec.\ \ref{sec:univ}, the FTG
function is found also for $\left({\cal R}{\rm e} c_{\nsd}\right)^2,
\left({\cal I}m c_{\nsd}\right)^2$, following derivations similar to
the ones in Sec.~\ref{sec:hm}.  So the FTG limits seem to be robust
with respect to the choice of expansion functions.  The FTG function
found for the case of hard modes (the equivalent of the result in
Sec.\ \ref{sec:hm}) explains the finding of \cite{PH}, where extremal
square-coefficient statistics was studied numerically for the XY
model, and the FTG function (\ref{Gumbel}) proved to be a good fit
when soft modes were discarded from the batches of intensities before
the selection of the maximal ones. 

\section{Comparison with the distribution of the roughness}
\label{sec:roughness}
 
Interestingly, there is a resemblance between the scaling function of
the extreme value PDFs for $\alpha>0$ and those for the roughness
\cite{falpha} for $\alpha>d$.  Namely, both types of PDFs have a
single maximum, positive skewness, a nonanalytic initial asymptote,
and a dominantly exponential decay (the leading term in the logarithm
of the PDF is linear for large $z$).  However similar they are
qualitatively, the functions are different.
 
A straightforward way to make the comparison quantitative is to
calculate the asymptotes of the PDFs.  For small $x$ the roughness PDF
behaves as given by (\ref{eq:E-Pd}) of Appendix \ref{sec:E}, and the
EVS PDF goes like (\ref{eq:logP}).   It is interesting to note that
for the roughness the critical $\alpha_c=d$, where the
variance shrinks to zero on the scale of the mean, while for the EVS
the same type of criticality is observed for $\alpha_c=0$.  Then both
asymptotes can be cast in the common form 
\begin{eqnarray}
  \label{eq:common}
  \ln P(y) \propto y^{-d/(\alpha-\alpha_c)},
\end{eqnarray}
where for the roughness and for EVS one should understand $y$ as $z$
and $x$, respectively.

Furthermore, in $1d$ one can calculate the power prefactor in front of
the nonanalytic exponential, {\em c. f.} (\ref{alphaasymp}) and
(\ref{eq:E-Pasymp}), that again can be written in the same form, so in
$1d$ both PDFs are asymptotically
\begin{eqnarray}
  \label{eq:common1d}
  P(y) \approx C_1 y^{-\frac{3(\alpha-\alpha_c)+2}{2(\alpha-\alpha_c)}} \,
  \times  \exp\left( - \frac{C_2}{y^{1/(\alpha-\alpha_c)}}\right),
\end{eqnarray} 
albeit the proportionality constants in the two asymptotes are
different.  What is more, in $1d$ both PDFs become the FTG function in
the limit $\alpha\to \alpha_c$.  Note however, that the two kinds of
PDFs have different asymptotes for larger arguments, when
$\alpha>\alpha_c$.

We can put the threshold behavior in a short form.  Firstly, for both
kinds of distributions there is a (lower) critical value
$(d/\alpha)_c^\ell$ where, for increasing $d/\alpha$, on the scale of
the mean the distributions become the Dirac delta, and this threshold
is $1$ for the roughness and $\infty$ for the EVS.  Then there is an
upper critical value $(d/\alpha)_c^u$, that is the threshold for the
respective classical limit distribution, Gaussian for the roughness
and FTG for the maximal intensity.  For the roughness we have
seen~\cite{falpha} that $(d/\alpha)_c^u=2$ and for the maximal
intensity it is again $(d/\alpha)_c^u=\infty$.  So there is a region,
$2\le d/\alpha \le \infty$, where the roughness is Gaussian, but the
EVS is still not given by any of the known limit distributions of EVS,
rather by the generalized Dedekind function. 

There is a significant difference also in the finite-size correction
to the PDF.  The PDF for the extremal amplitude converges essentially
exponentially fast (see App.~\ref{sec:FS} for the $1d$ correction
formula), while the correction to the roughness PDF can be shown to be
algebraic in $N$.
 
In sum, there is a strong qualitative resemblance between the shape of
the roughness PDF for $\alpha>d$ and the EVS, furthermore, their
initial asymptotes have similar functional forms, they are, however,
distinctly different functions.  It remains to be clarified whether
the similarity has some deeper reason, or it is simply a mathematical
coincidence.
  
\section{Conclusion}  
\label{sec:concl}  
 
The results presented above bring us to a very definite conclusion
concerning the relation between interface fluctuations and extremal
statistics: the roughness PDF is not given by the largest mode
in the Gaussian interface model.  This is true, even in the case
$\alpha=d=1$, where the roughness PDF is the FTG
distribution~\cite{1f}, one of the known limit functions of extreme
statistics.  We showed, further, that the PDF was none of the known
EVS limit distributions and depended on the statistics for the
individual elements of the model.  In addition, one expects the
boundary conditions to influence the shape of the PDFs.  It should be
added that, to the extent that the Gaussian model is a universal
family of massless models from the viewpoint of critical phenomena,
the generalized Dedekind PDFs are equally universal, depending on the
dimensionality $d$ and the dispersion parameter $\alpha$. 
 
It is worth pointing out that there is an analogy between the Gaussian
distribution arising from the central limit theorem, which applies for
sums of random variables with finite moments, and the extreme value
limit distributions such as FTG, which is about the maximal of those
variables.  Both limit distributions are related to a large ensemble
of independent, identically distributed objects.  The appearance of a
non-Gaussian PDF for the integrated power spectrum ( i.e.\ roughness),
in Gaussian systems
\cite{BHP98,Pinton2,BHP2000,Bram01,PH,FORWZ,RP94,1f,falpha}, is a
consequence of strong dispersion, whence follows the strongly
non-identical distribution of the modes.  What we have illustrated in
this paper is that it is this same dispersion that generically
excludes the known limit distributions of extreme statistics.  To
refine the picture, we found that if Gaussian central limit statistics
is excluded for the integrated variable because of dispersion, FTG
extremal statistics is also explicitly excluded, but the reverse is
not necessarily true: there is the region of finite $d/\alpha\ge 2$,
where the integrated power is Gaussian but the maximal Fourier
intensity does not follow FTG.  However, we found three border cases
where, by singular scaling, the FTG distribution arises.  In the limit
$\alpha\to 0$ the amplitudes obviously become identically distributed.
For the other cases, that is, in the limit $d\to\infty$ and when a
large number of modes are omitted near the center of the Brillouin
zone, the number of contributing modes diverge but the dispersion
across the zone remains.  This result appears to broaden the scope of
validity of the FTG distribution in EVS to non-i.\ i.\ d.\ variables.
Furthermore, the special scalings that led to the emergence of FTG in
the border cases demonstrate new ways of extracting the known limit
distribution even in the absence of the conventional FTG scenario.
 
In summary, the extreme value PDF for Gaussian interface models are,
for fixed choice of variables, a two-parameter family of functions
with a nonanalytic part for small values, a single maximum, and an
essentially exponential tail. The functions look qualitatively
similar, but vary quantitatively over the range of parameters $d$ and
$\alpha$ studied.  This family of curves is different from that found
for the roughness distribution for the same parameters $\alpha,d$
\cite{falpha}. We conclude that it is not possible to make a direct
link between non-Gaussian roughness fluctuations and extreme values in
these models. Given the multitude of recent observations of
non-Gaussian fluctuations in experimental and in model strongly
correlated systems, as well as in the use of interface models as
phenomenological tools in describing such fluctuations, this seems
like an important result.  However, the relevance of extreme values in
more complex, non-Gaussian systems remains an open question and it
would be interesting to follow up this work with studies in this
direction.
 
\acknowledgments
 
This work was supported, the by the Hungarian Academy of Sciences
(Grant Nos. OTKA T029792 and T043734) and by CNRS ACI grant no. 2226.
One of us (ZR) thanks the Ecole Normale Sup\'erieure for financial
support during his visit to Lyon.
 

 
\appendix

\section{Abbreviations} \label{sec:Abbr}

Below we summarize the abbreviations used throughout the paper.

\vspace{\baselineskip}
\noindent
extreme value statistics  \dotfill EVS  \\
independent, identically distributed \dotfill i.\ i.\ d. \\
probability density function  \dotfill PDF  \\
integrated probability distribution function \dotfill IPDF\\
Fisher-Tippett-Gumbel \dotfill FTG\\
Fisher-Tippett-Fr\'echet \dotfill FTF\\

\section{Small-$z$ asymptote of the extreme value distribution for the
  intensities} \label{sec:A}
\subsection{ $d=1$}  
\label{sec:A1}  
 
Our starting formula is Eq.\ (\ref{ipdf1}) where, for small $z$, we
expand terms with not too large $n$ to leading order in $n^\alpha z$.
Terms with larger $n$s must be considered without expansion.  However,
exponentiating them the product becomes a sum of logarithms, which we
can replace by an integral as the terms vary slowly with $n$.
Carefully treating various corrections allows us to obtain the
asymptote for (\ref{ipdf1}).
 
Let us separate the product (\ref{ipdf1}) for the IPDF, while $z\to
0$, as
\begin{equation}
  \label{eq:A-sep}
M_\alpha(z) =  C \times D,
\end{equation}
where
\begin{eqnarray}
C &=& \prod_{n=1}^{n_\alpha} \left( 1-e^{-n^\alpha z} \right) \approx
\prod_{n=1}^{n_\alpha} n^\alpha z = z^{n_\alpha} \left(n_\alpha
  !\right)^\alpha, \label{eq:A-A} \\
D &=& \exp \left(\sum_{n=n_\alpha+1}^\infty \ln\left( 1-\exp(-n^\alpha
    z)\right)\right),
\label{eq:A-B}
\end{eqnarray}
with $(n_\alpha)^\alpha z$ small but $n_\alpha$ large, i.\ e.\ 
\begin{eqnarray}
  \label{eq:A-ineq}
\frac{1}{z^{1/\alpha}} \gg n_\alpha \gg 1, 
\end{eqnarray}
The first inequality allows the linearization of the exponential in
(\ref{eq:A-A}), whereas the second one will enable us to use the
Stirling formula in (\ref{eq:A-A}) and replace the sum in
(\ref{eq:A-B}) by an integral.
 
A short detour is necessary to see under what condition the correction
to the term linear in $z$ in each factor of (\ref{eq:A-A}) can be
neglected. Including the next term in the expansion we have
\begin{eqnarray}
C& \approx& \prod_{n=1}^{n_\alpha} n^\alpha z  \left(
  1- \frac{1}{2} n^\alpha z \right)  \nonumber   \\
& \approx&   \exp\left(\frac{z}{2} \sum_{n=1}^{n_\alpha} n^\alpha
  \right)  \prod_{n=1}^{n_\alpha}n^\alpha z.
\label{eq:A-A-corr} \end{eqnarray} 
The exponent in the prefactor goes like $z \,n_\alpha^{\alpha+1}$.  So
the prefactor can be taken as unity if we use an $n_\alpha$ that
satisfies
\begin{eqnarray} 
\frac{1}{z^{1/(\alpha+1)}} \gg n_\alpha \gg 1, 
\label{eq:A-eps} \end{eqnarray} 
a condition stricter than (\ref{eq:A-ineq}).  Then the approximation
in (\ref{eq:A-A}) indeed gives the asymptote of $C$.
 
Now we can calculate $C$ from the r.\ h.\ s.\ of (\ref{eq:A-A}) by
using Stirling's formula
\begin{eqnarray}
C &\approx& z^{n_\alpha}  \,  \left( \frac{n_\alpha}{e}\right) ^{\alpha
  n_\alpha}   \left(2\pi n_\alpha \right) ^{\alpha/2}.
\label{eq:A-A-appr} \end{eqnarray}

To calculate $D$ we first estimate the error
incurred when the sum is written as an integral: given
a function $f(z)$, the terms in the sum of $f(n)$ from $n_\alpha+1$ to
infinity can be written to leading order
\begin{eqnarray}
f(n+1) \approx \int_n^{n+1} \!\! \, \td x\, f(x) + \frac{1}{2} f^\prime (n+1).
\label{eq:A-B-sum2int-1} \end{eqnarray}
Performing the summation, we can replace the sum of the
$f^\prime(n+1)$s by an integral in leading order and we finally find
up to the next-to-leading order
\begin{eqnarray}
\sum_{n=n_\alpha}^\infty f(n+1) \approx \int_{n_\alpha}^{\infty} \!\!
 \, \td x\, f(x) -  \frac{1}{2} f(n_\alpha).
\label{eq:A-B-sum2int-2} \end{eqnarray} 
This is the exponent of (\ref{eq:A-B}) with 
\begin{eqnarray}
\label{eq:A-B-f} 
f(x)=\ln\left(1-\exp(-x^\alpha z)\right) 
\end{eqnarray} 
whose integral can be written using $u=x\,z^{1/\alpha}$ as
\begin{eqnarray}
 \int_{n_\alpha}^{\infty} \!\! \, dx\, f(x)  &=& \frac{1}{z^{1/\alpha}}
  \int_{n_\alpha z^{1/\alpha}}^{\infty} \!\! \, du\, \ln\left(
  1-\exp(-u^\alpha ) \right)\nonumber \\
  &\approx & \frac{1}{z^{1/\alpha}}
  \int_{0}^{\infty} \!\! \, du\, \ln\left(  1-\exp(-u^\alpha ) \right)
  \nonumber \\ && - \frac{\alpha}{z^{1/\alpha}}
  \int_0^{n_\alpha z^{1/\alpha}} \!\! \, du\, \ln u.
\label{eq:A-B-int} \end{eqnarray}
In the last line we used the smallness of $u$: one can easily convince
oneself, from (\ref{eq:A-eps}) that the higher order terms
vanish.  The definite integral is
\begin{eqnarray}
c(\alpha) &=& -\int_{0}^{\infty} \!\! \, \td u\,    \ln\left(
  1-\exp(-u^\alpha ) \right)\nonumber \\  &= &
  \zeta\left(1+\frac{1}{\alpha}\right)
  \, \Gamma\left(1+\frac{1}{\alpha}\right),
\label{eq:A-c}
 \end{eqnarray}
and so from
 (\ref{eq:A-B-sum2int-2}) we get
\begin{eqnarray}
\ln D \approx -\frac{c(\alpha)}{z^{1/\alpha}} + \alpha n_\alpha -
\left( n_\alpha + \frac{1}{2}\right)  \ln\left((n_\alpha)^\alpha z\right).
\label{eq:A-B-final}
 \end{eqnarray}
Finally,
 (\ref{eq:A-A-appr}) and (\ref{eq:A-B-final}) give
\begin{eqnarray}
M_\alpha(z) \approx \frac{(2\pi)^{\alpha/2}}{\sqrt{z}} \exp \left(
  -\frac{c(\alpha)}{z^{1/\alpha}}\right).
\label{eq:A-M-final}
 \end{eqnarray}
 This is the sought after asymptote for small $z$ in the sense that
its relative error vanishes for $z\to 0$.
 
Remarkably, when we
neglect the large $n$ contribution, we find the correct
$1/z^{1/\alpha}$ dependence in the argument of the exponential in
${M}_\alpha(z)$.  However, the coefficient $c(\alpha)$ is only found by
keeping terms with large $n$.  Note, furthermore, that the inequality
(\ref{eq:A-eps}) needs $\alpha>0$.  Thus we should not be surprised
that the asymptote for $\alpha\to 0$ does not relate to the tail of
the FTG function (\ref{Gumbel}) for large negative argument.
 
In summary, the leading term of $\ln M(z)$ was produced by
exponentiating the product and rewriting the sum as an integral.  We
shall follow this recipe below for general dimensions. 

\subsection{Arbitrary  dimension}  \label{sec:B1}
 
We consider here the asymptotes of the IPDF (\ref{eq:B-M}).  The
leading term of the logarithm of the IPDF $M_{\alpha,d}(z)$, in the
small-$z$ limit, can be determined by transforming the sum over the
modes of the Brillouin zone into an integral, because for $z\to 0$
the terms in the sum change slowly.  Thus we obtain, to leading order
in $z$
\begin{eqnarray} 
  \label{eq:B-logM}  
\ln {M}_{\alpha,d}(z) \approx - \frac{c(\alpha,d)}{z^{d/\alpha}},
\end{eqnarray}
where
\begin{eqnarray} 
c(\alpha,d) &=& -\frac{1}{2} \int\td^d u \, \ln \left( 1 - \exp\left( -  | {\bf
 u}|^\alpha\right) \right) \nonumber \\
 &=&   \frac{\pi^{d/2} \zeta\left(1+\frac{d}{\alpha}\right) \,
 \Gamma\left(1+\frac{d}{\alpha}\right)}{d\, 
 \Gamma\left(\frac{d}{2}\right)}.  \label{eq:B-c} 
\end{eqnarray}
For $d=1$ we indeed recover the $c(\alpha)$ of (\ref{eq:A-c}).  Note
that (\ref{eq:B-logM}) does not give the full asymptote for $M$, it is
only the leading term in the exponent.
 
In order to estimate the next-to-leading term for $\ln
M_{\alpha,d}(z)$ one can calculate the correction arising when the sum
is transformed into an integral.  This is of the order
\begin{eqnarray}
  \label{eq:B-logM-err}
 \frac{\ln z }{z^{(d-1)/\alpha}}.
\end{eqnarray}
This correction diverges algebraically for $d>1$, giving rise to a
further exponential singularity in the IPDF. The asymptote of $M$ is
therefore not as simple as in the $1d$ case (\ref{eq:A-M-final}).

\section{Small-$z$ asymptote in the case of  the sine and cosine expansion}
\label{sec:A2}

In this Appendix we just summarize the results for the PDF of the
maximal square-coefficient for the sine and cosine expansion
functions.  The leading term for the logarithm of the IPDF
(\ref{eq:Mgenerald2}) comes from our making the sum an integral
\begin{eqnarray} 
  \label{eq:B-logM2} 
\ln {\tilde M}_{\alpha,d}(z) \approx - \frac{c(\alpha,d)}{z^{d/\alpha}},
\end{eqnarray} 
where 
\begin{eqnarray}  
{\tilde c}(\alpha,d) = - \int \td^du \,\ln \left( {\rm erf}\left(
  |{\ud}|^{\alpha/2} 
  \right) \right).
\label{eq:B-c2}
\end{eqnarray} 
In $d=1$, after a calculation along a line similar to the one followed
in Appendix \ref{sec:A1}, we obtain the full asymptote
\begin{eqnarray} 
{\tilde M}_{\alpha,1}(z) \approx
\frac{(2\pi)^{\alpha/2}\sqrt{\pi}}{2\sqrt{z}} \exp \left(
  -\frac{{\tilde c}(\alpha,1)}{z^{1/\alpha}}\right).
\label{eq:A-M2-final}  
\end{eqnarray}
Note that the functional form is similar to the asymptote of the IPDF
for the maximal intensity (\ref{eq:A-M-final}), but the constants are
different.

\section{Finite-$N$ correction} \label{sec:FS} 
 
Here we calculate the correction of the extreme intensity distribution
for large $N$.  Denoting now the finite product (\ref{eq:B-M}) by
$M_{N,\alpha,d}(z)$ and assuming $N$ to be large we obtain
\begin{eqnarray} 
  \label{eq:FS-1}
  &&\frac{M_{N,\alpha,d}(z)}{M_{\alpha,d}} =
  \prod_{|{\nsd}|>N}\!\!\!\pr \left( 1-e^{-|{\nsd}|^\alpha
  z}\right) \nonumber \\
  &&\approx 1 - \frac{1}{2} \sum_{|{\nsd}|>N} e^{-|{\nsd}|^\alpha
  z} = 1-I_{N,\alpha,d}(z). 
\end{eqnarray} 
Hence 
\begin{eqnarray}
  \label{eq:FS-IPDF} 
   M_{N,\alpha,d}(z) &\approx& M_{\alpha,d}(z) \left( 1+
   I_{N,\alpha,d}(z) \right), \\
  \label{eq:FS-PDF}
   P_{N,\alpha,d}(z) & \approx & P_{\alpha,d}(z) +
    M_{\alpha,d}(z)\, I^\prime_{N,\alpha,d}(z),
\end{eqnarray} 
where the second line was obtained by differentiation and we used the
property that $I$ is negligible next to $I^\prime$ for large $N$.
From (\ref{eq:FS-1}) we surmise that the convergence to the $N=\infty$
functions is in essence exponentially fast in $N^\alpha$; to be
specific, we give below for $1d$ the precise asymptote.
 
Considering in $d=1$ the $\alpha=1$ case, we have a geometric series
as
\begin{eqnarray} 
   \label{eq:FS-a1d1}
  I_{N,1,1}(z) = \sum_{n=N+1}^\infty e^{-n z}  =
  e^{-Nz}\,\left(e^{z}-1\right)^{-1}. 
\end{eqnarray}
If $\alpha>1$ then
\begin{eqnarray}
  \label{eq:FS-ald1}
  I_{N,\alpha,1}(z) &=& \sum_{n=N+1}^\infty e^{-n^\alpha z} =
  \sum_{n=N+1}^\infty e^{-n\,n^{\alpha-1} z} \nonumber \\
   &<& \sum_{n=N+1}^\infty e^{-n\,(N+1)^{\alpha-1} z} = \frac{
  e^{-(N+1)^{\alpha} z}}{1- e^{-(N+1)^{\alpha-1} z}} \nonumber \\
  &\approx & e^{-(N+1)^{\alpha} z}.
\end{eqnarray} 
The last formula, an upper bound for the asymptote, is just the first
term in the sum $I$.  The sum has positive terms, whence it follows
that the first term is at the same time the asymptote of $I$ for
$\alpha>1$.  Finally, in the case $\alpha<1$ we can rewrite the sum
for $I$ into an integral, because if in Eq.~(\ref{eq:A-B-sum2int-1})
we substitute $e^{-n^\alpha z}$ for $f(n)$ and $N+1$ for $n_\alpha$,
the correction to the integral is negligible for $\alpha>1$ in the
large $N$ limit.  The integral is approximately $N^{1-\alpha}
e^{-N^{\alpha}}/\alpha\,z$, this is then the sought asymptote for $I$.
In conclusion,
\begin{eqnarray} 
  \label{eq:FS-agd1}  
  I_{N,\alpha,1}(z) 
   \left\{ \begin{array}{lr} 
        \approx e^{-(N+1)^{\alpha} z},& \mathrm{if~} \alpha>1, \\ 
        = e^{-Nz}\,\left(e^{z}-1\right)^{-1},&  \mathrm{if~} \alpha=1,  \\
       \approx N^{1-\alpha} e^{-N^{\alpha} z} / \alpha\,z,&
   \mathrm{if~} \alpha<1. \end{array} \right. 
\end{eqnarray}
We can summarize the above results for various $\alpha$s such that if
the summand in $I$ decays slowly, one can replace the sum by an
integral, and if it decays fast then the sum asymptotically equals its
first term. 
 
\section{On the FTG limits} \label{sec:C} 
 
Below we derive Eq.\ (\ref{eq:a0}) from (\ref{eq:a0M1}). We shall
replace the lower limit of integration $1$ by $n_0$ to show
irrelevance of the precise setting of the lower limit.  Denoting the
integral in (\ref{eq:a0M1}) by $I$ and changing the integration
variable to $v=n^\alpha z$ we get
\begin{eqnarray} 
I &=& \frac{1}{\alpha\,z^{d/\alpha}}\, \int_{n_0^\alpha z}^\infty e^{-v}\,
  v^{d/\alpha-1}\td v. 
\label{eq:C1} 
\end{eqnarray} 
Since $\alpha\to 0$, the lower integration limit becomes $z$, indeed
independent of $n_0$.  The saddle point method gives
\begin{eqnarray} 
I &\approx& \sqrt{\frac{\pi}{2\alpha d}} \, \left(
  \frac{d}{z\alpha e} \right)^{d/\alpha}\, {\rm erfc}
  \left(\frac{z\alpha-d}{\sqrt{\alpha d}}\right).
\label{eq:C2} 
\end{eqnarray} 
When $z \approx d/\alpha e$, the argument of the erfc becomes a large
negative number, where the erfc is approximately 2.  Thus, including
the prefactor before $I$ in (\ref{eq:a0M1}), we recover 
(\ref{eq:a0}).  
 
We have here an opportunity to test whether the sum in (\ref{eq:a0M1})
was justly rewritten into an integral.  We do not have rigorous
results, but we know that the integral must not be smaller than a
single summand term, if it is, the integral representation cannot be
accepted.  A characteristic summand term now is $e^{-n^\alpha
  z}\approx e^{-v}$, where $v = (d-1)/\alpha$ is the saddle point in
(\ref{eq:C1}), so the summand term is small.  Since we consider $z$s
for which the full integral for $-\ln M$ is of $O(1)$, the integral is
indeed much larger then a single characteristic summand term $e^{-v}$.
This is a generic test that the integral representation of a sum
should pass, and we performed it in all pertinent cases in the paper.

\section{Small-$z$ asymptote of the roughness distribution} \label{sec:E}  
\subsection{One dimension, $\alpha>1$} \label{sec:E-1} 
 
We derive here the small-$x$ asymptote of the PDF of the roughness for
the Gaussian model of $1/f^\alpha$ noise with periodic boundary
condition.  We consider the case $\alpha> 1$, when all cumulants of
the roughness are of the same order.  In the case of the Wiener
process, $\alpha=2$, the PDF has a nonanalytic asymptote, which has
been calculated in Ref.~\cite{FORWZ}.  Here we find that
nonanalyticity prevails for all $\alpha> 1$.  We begin with the
simplest form, free of normalizing constants, of the generating
function~\cite{falpha}
\begin{eqnarray} 
  \label{eq:E-G} 
G_{\alpha}(s) =  \prod_{n=1}^\infty  \left(1+\frac{s}{n^\alpha}\right)^{-1}. 
\end{eqnarray} 
The average is given by $-G_{\alpha}^\prime(0)=\zeta(\alpha)$, where
$\zeta$ is Riemann's zeta function, so in order to obtain a PDF
normalized to unit average, in the end we should rescale by
$\zeta(\alpha)$.

The large-$s$ asymptote of the generating function will give us the
initial asymptote of the PDF. The calculation goes along the lines of Appendix
\ref{sec:A}.  For large $s$ we can factorize $G_{\alpha}$ as
\begin{equation}
  \label{eq:E-sep}
G_\alpha(s) =  E \times F,
\end{equation}
where
\begin{eqnarray}
E &=& \prod_{n=1}^{n_\alpha} \left( 1+\frac{s}{n^\alpha} \right)^{-1} \approx
\prod_{n=1}^{n_\alpha} \frac{n^\alpha}{s} =
s^{-n_\alpha}\left(n_\alpha !\right)^\alpha, \label{eq:E-C} \\
F &=& \exp \left(-\sum_{n=n_\alpha+1}^\infty \ln\left(
    1+\frac{s}{n^\alpha}\right)\right), 
\label{eq:E-D}
\end{eqnarray}
where
\begin{eqnarray} 
  \label{eq:E-ineq} 
s^{1/(\alpha+1)} \gg n_\alpha  \gg 1.
\end{eqnarray}
The first inequality ensures that (\ref{eq:E-C}) indeed gives the
asymptote for $E$, see App.\ \ref{sec:A} for an analogous estimate,
and the largeness of $n_\alpha$ enables us to approximate the sum in
(\ref{eq:E-D}) by an integral.  We also calculate in $F$ the
nonvanishing corrections to the integral, again in a way similar to
what was done in App.\ \ref{sec:A}.  Then using the Stirling formula
in (\ref{eq:E-C}) for $E$, substituting the $E$ and $F$ into
(\ref{eq:E-sep}), and collecting all stray terms we wind up for large
$s$ with
\begin{eqnarray}
  \label{eq:E-Gfinal}
  G_\alpha(s) &\approx& \left( 2\pi\right)^{\alpha/2} \sqrt{s} \,
  \exp\left(-s^{1/\alpha}\, g(\alpha)\right),
\end{eqnarray} 
where
\begin{eqnarray}
\label{eq:E-g}
  g(\alpha) &=& \int_0^\infty \!\! \td u \,
  \ln\left(1+u^{-\alpha}\right) =
  \frac{\pi}{\sin\left(\frac{\pi}{\alpha}\right)}.
\end{eqnarray} 
The PDF of the roughness, normalized to unit mean, is obtained by
inverse Laplace transformation
\begin{eqnarray} 
  \label{eq:E-Laplace} 
  P_\alpha(x) = \int \frac{\td s}{2\pi i} e^{sx}\, 
  G_\alpha\left(\frac{s}{\zeta(\alpha)}\right).  
\end{eqnarray} 
For small $x$ the large-real-$s$ region dominates, where the saddle point
method can be used.  It suffices to compute the saddle point from the
exponential of (\ref{eq:E-Gfinal}), and then substitute its value into
the $\sqrt{s}$ prefactor.  Taking into account also the quadratic
deviation from the saddle point in the exponent, we finally obtain the
small-$x$ asymptote 
\begin{eqnarray} 
  \label{eq:E-Pasymp} 
  P_\alpha(x) \approx Q(\alpha)\, x^{-\frac{3\alpha-1}{2(\alpha-1)}}\,
  \exp\left(-\frac{R(\alpha)}{x^{1/(\alpha-1)}}\right), 
\end{eqnarray} 
where 
\begin{eqnarray}
  \label{eq:E-Q} 
  Q(\alpha)& = &  \frac{(2\pi)^\frac{\alpha-1}{2} g(\alpha)^\frac{\alpha}{\alpha-1}
  } {\sqrt{\alpha-1}\,
  \left(\alpha\zeta(\alpha)\right)^\frac{\alpha+1}{2(\alpha-1)}}, \\ 
  \label{eq:E-R} 
  R(\alpha)& = &  \frac{(\alpha-1)\,g(\alpha)^\frac{\alpha}{\alpha-1}}
  {\alpha^\frac{\alpha}{\alpha-1}\, \zeta(\alpha)^\frac{1}{\alpha-1}}. 
\end{eqnarray} 

\subsection{Arbitrary dimension, $\alpha>d$} \label{sec:E-2}

In general dimensions we can give here only the leading exponential
for the initial asymptote of the PDF of the roughness for periodic
boundary condition.  We consider the case $\alpha>d$, else there is no
sense in speaking about the asymptote for small roughness on the scale
$L^{d-\alpha}$, see Ref.~\cite{falpha}.  The generating function
yielding unit average is from~\cite{falpha}
\begin{eqnarray} 
  \label{eq:E-Gd} 
G_{\alpha,d}(s) &= & \prod_{|\nsd|>0} 
\left(1+\frac{s}{\zeta(\alpha,d)\,|\nd|^\alpha}\right)^{-1/2},
\end{eqnarray} 
where
\begin{eqnarray}  
  \label{eq:E-zd}  
\zeta(\alpha,d)= \frac{1}{2} \sum_{|\nsd|>0}|\nd|^{-\alpha}
\end{eqnarray} 
is a $d$-dimensional generalization of the Zeta function.
Exponentiation of the product to a sum and transformation of the sum
to an integral gives the leading exponential term
\begin{eqnarray}   \label{eq:E-Gfinald}
\ln G_{\alpha,d}(s)\approx
-\left(\frac{s}{\zeta(\alpha,d)}\right)^{d/\alpha}\,  g(\alpha,d), 
\end{eqnarray} 
where
\begin{eqnarray} 
  g(\alpha,d) &=& \int_0^\infty \!\! \td^d u \,
  \ln\left(1+|\ud|^{-\alpha}\right) \nonumber \\ &=&
  \frac{\pi^{d/2+1}}{d\, \Gamma\left(\frac{d}{2}\right)\,\sin\left(\frac{\pi
  d}{\alpha}\right)}.   \label{eq:E-gd}  
\end{eqnarray}  
Performing the inverse Laplace transformation by the saddle point
method, one arrives at the small-$x$ asymptote of the PDF as
\begin{eqnarray}   \label{eq:E-Pd}
\ln P_{\alpha,d}(x)\approx
-\frac{R(\alpha,d)}{x^\frac{d}{\alpha-d}},
\end{eqnarray} 
with
\begin{eqnarray}   \label{eq:E-Rd} 
R(\alpha,d) = \frac{\alpha-d}{d}\, \left[
  \frac{\pi^{d/2+1}}{\alpha\, \Gamma\left(\frac{d}{2}\right)\,
  \sin\left(\frac{\pi d}{\alpha}\right)\, \zeta(\alpha,d)^{d/\alpha}} \right]
^\frac{\alpha}{\alpha-d}\!\!\!\!\!\!\!\!\!\!\!.
\end{eqnarray}




\begin{thebibliography}{38}
\bibitem[{Not({\natexlab{a}})}]{NoteEVS}
\bibinfo{note}{In the statistical physics literature the limit functions are
  often called universal, but we do not use this term here in order to avoid
  confusion with universality in critical systems.}
\bibitem[{Not({\natexlab{b}})}]{NoteFTG}
\bibinfo{note}{The FTG IPDF is often written as $M_{\mathrm{
  FTG}}(y)=e^{-e^{-y}}$, the present form (\ref{Gumbel-IPDF}), however, has a
  natural scale for the extremal variable. Namely, in order to test whether an
  empirical extremal quantity $z$ is of the FTG class, one can compute the
  variable $y=(z-\left<z\right>)/\Delta z$, where $\left<z\right>$ and $\Delta
  z$ are the mean and the standard deviation of $z$, respectively, and compare
  the distribution of $y$ to (\ref{allGumbel}).}
\bibitem[{par()}]{parent}
\bibinfo{note}{Traditionally, the term parent PDF refers to the single
  distribution function of i.\ i.\ d.\ variables, but we shall speak about
  parent PDFs also in the case, when the variables are independent, but
  differently distributed.}
\end{thebibliography}
\end{document}